\documentclass{article}
\usepackage{amsmath}
\usepackage{amssymb}
\usepackage[utf8]{inputenc}
\usepackage{graphicx}
\usepackage{hyperref}
\usepackage{latexsym}
\usepackage{setspace,bm} 

\renewcommand{\d}{\partial}

\newcommand{\ket}[1]{\left|{#1}\right\rangle}
\newcommand{\Tr}{\mathop{\textrm{Tr}}}
\newcommand{\nn}{\nonumber\\}

\begin{document}

\title{Spontaneous Symmetry Breaking without classical fields: a Functional Renormalization Group approach}
\author{A. Jakov\'ac, P. Mati, P. P\'osfay\\
\small\it Department of Computational Sciences, Wigner Research Centre for Physics,\\\small\it29-33 Konkoly-Thege Mikl\'os Street, H-1121 Budapest, Hungary}
\date{\today}
\maketitle
\begin{abstract}
    We propose an approach to describe Spontaneous Symmetry Breaking (SSB) that does not rely on the order parameter dependent free energy (Landau theory). We use the Functional Renormalization Group (FRG) evolution of the explicitly broken theory, using a truncation scheme that is compatible with the Ward identities. To represent the symmetry breaking, we propose to use the "Ward ratio" which is zero in the symmetric phase and unity in the broken phase. In this approach a unified scale evolution of the effective potential is applicable in both phases. It is peculiar that the scale evolution is accelerated in the critical regime.
\end{abstract}

\section{Introduction}

We try to overview the problem of Spontaneous Symmetry Breaking (SSB) from the point of view of Functional Renormalization Group (FRG). SSB is one of the most successful proposals in many particle quantum mechanics and quantum field theory (QFT). With the help of this concept we could give an account for superconductivity \cite{10.21468/SciPostPhysLectNotes.11}, superfluidity \cite{PhysRevA.81.053630}, Higgs mechanism and weak gauge boson masses \cite{Bhattacharyya_2011}, and we could continue the list.

Usually we call a transformation symmetry, if under its action the dynamics (for example the Hamiltonian) is invariant \cite{10.21468/SciPostPhysLectNotes.11}. It is still possible that the symmetry is not manifested in the observables, then we speak about spontaneous symmetry breaking (SSB).

To describe SSB, we may apply the Landau theory of phase transitions \cite{HOHENBERG20151}. We shall define the order parameter of the symmetry, which is a microscopic observable that transforms with an irreducible representation under the symmetry transformation. We shall calculate the free energy (effective potential) at a fixed value of the order parameter. If the minimum of the effective potential is at an order parameter value which is invariant under the symmetry transformation, then we are in the symmetric phase, otherwise we are in the broken phase. In the broken phase there must be multiple minima of the free energy, connected by the symmetry transformation.

To implement Landau theory in quantum systems \cite{10.21468/SciPostPhysLectNotes.11}, we need an order parameter operator whose ground state (vacuum) expectation value yields the value of the order parameter. The order parameter operator transforms with an irreducible representation of the symmetry transformation. But its vacuum expectation value is still zero, if the vacuum itself is invariant. Thus the order parameter may have nonzero expectation value, if both the order parameter operator and also the vacuum itself are not invariant under the symmetry transformation. As it is usual to express, in SSB the dynamics reflects the symmetry, but the ground state (the vacuum) breaks it. As an immediate consequence, there must be several vacua that are connected by the symmetry transformation.

The concept of the order parameter, and the order parameter dependent effective potential, however, sometimes lead to spurious consequences. Let us recall some of the challenges. 
\begin{enumerate}
    \item In Quantum Mechanics we have a theorem that the ground state is unique, thus it must be a singlet state, while in SSB the vacuum must change under the symmetry transformation. This seems to be in contradiction with the several vacua required by the SSB. The only way out \cite{Liu:2005ofe,LANDSMAN2013379} is that there are superselection classes with ground states $\ket a$ where these states are unitary inequivalent (i.e. there is no physical (local) unitary transformation $U$ that would have nonzero matrix elements between these vacua). The drawback of this solution is that, although it makes it possible to have equal energy multiple vacua, but the lack of communication forbids the change between them by physical processes, even in the presence of external fields. This means that we are stuck to a single vacuum forever; this would forbid for example the phenomenon of hysteresis. If, on the other hand the transition matrix elements are not zero, then a new singlet ground state is formed where the symmetry is not broken. Actually, this is the problem of the quantum computers: maintaining a state against decoherence effects, and the need to change the states in order to perform computations are two contradicting requirements.
    \item The effective potential, which is the effective action for constant order parameter, is the result of a Legendre transformation, and therefore it must be convex \cite{touchette2005legendre}. If there are several minima with the same energy, then convexity requires flat potential for a certain domain. But the derivatives of the effective potential at the minimum correspond to physical quantities, for example the second derivative gives the particle mass. A flat potential would yield massless and free particles, which is clearly not correct. Even in the presence of a small external field which explicitly breaks symmetry, the curvature mass remains close to zero, as opposed to the observations.
    \item It is not clear that the effective potential is sensible at all anywhere else than in the infinitesimal open neighborhood of the vacuum. In gauge theories, for example, the effective potential and its derivatives are gauge invariant only at the physical point \cite{PhysRevD.9.1686,Andreassen_2015}.
\end{enumerate}

Although there are practical recipes to overcome these difficulties, but they still exhibit a motivation to rephrase the mechanism of SSB in a different language. In this rephrasing we shall avoid all of the problems listed above.

To avoid problem one, we shall use a single vacuum, which is the unique ground state of the system. To comply with the problems of two and three, we shall use only the infinitesimal neighborhood of the vacuum, where the effective potential is striclty convex. Since we do not try to include those regions into the domain where the effective potential would be concave, no flattening will take place.

The price we have to pay is that in the vicinity of the vacuum the effective potential is not manifestly symmetric in the case of SSB. The remnants of the symmetry are the Ward identities. This suggests that we shall use a description that can give an account for the Ward identities \cite{Gies_2012}.

The other difficulty is that we lose the direct handle to the background field. We can approach it indirectly, for example through the non-symmetric couplings or through the Ward identities.

In this paper the above program is performed on the simplest case, the $\mathbb{Z}_2$ symmetric quartic scalar model. In order to describe the theory around the vacuum both in the symmetric and the broken phase, we need to introduce also symmetry breaking terms. Up to $\Phi^4$ power, the only candidate is the cubic interaction.

The technique to obtain the effective action will be the Functional Renormalization Group (FRG) \cite{Gies_2012}, in the LPA approximation. We will follow the scale evolution of the couplings of the theory written up around the physical vacuum, containing also symmetry breaking terms. We also have to give an account to the changing vacuum, this will result a small difference in the FRG equations \cite{Kobayashi_2020}.

The paper will be organized as follows. In Section \ref{sec:FRG} we overview the basic definitions of the regularization dependent effective action, and its FRG evolution equation. In Section \ref{sec:vacuumexp} we demonstrate the method, how can one write up the FRG evolution equation for a system with a changing vacuum. The same computation with the standard techniques is repeated in Section \ref{sec:vacuumexp_alt} to convince the reader about the correctness of the equations. In Section \ref{sec:results} we analyze the typical behaviour of the running, and demonstrate that the structure of the Ward ratio indeed reflects the symmetry group of the action. Section \ref{sec:discussion} is devoted to the discussion of the results, for the outlook and conclusions.

\section{Functional Renormalization Group}
\label{sec:FRG}

In quantum field theory or in many particle quantum theory, in order to calculate the effective action, we can use the method where the quantum fluctuations are introduced gradually. To this end we add a quadratic "regulator" term $R_k$ to the Lagrangian, where $k$ corresponds to some scale, and compute the partition function
\begin{equation}
    \label{eq:regZk}
    Z_k[J] = \int{\cal D}\Phi e^{-S[\Phi]-\frac12 \int \Phi R_k\Phi + \int J\Phi}.
\end{equation}
The regulator should satisfy in Fourier space
\begin{equation}
    R_k(p) = \left\{
    \begin{array}{cc}
         0 &\mathrm{for}\; p\gg k \\
         \mathrm{large} &\;\mathrm{for}\; k\gg p.
    \end{array}
    \right.
\end{equation}
This means that small momentum fluctuations are suppressed because they have a large effective mass. By lowering $k$ more and more low energy modes are taken into account. As we see, $Z_k[J]$ is a result of a quantum field theory with a modified action. 

From $Z_k[J]$ we compute the effective action using the Legendre transformation
\begin{equation}
    \Gamma_k[\Phi] + \frac12 \int \Phi R_k\Phi = \sup_{J\in D}\left(\int J\Phi - \ln Z_k[J]\right),
\end{equation}
where $D$ is the domain of the $J$ variable. We shall remark that the Legendre transformation depends on the choice of the domain. If we choose $D$ to be an infinitesimal neighborhood of the $J=0$ physical point, then $\Gamma_k[\Phi]$ will be analytic and convex in the infinitesimal neighborhood of
\begin{equation}
    \Phi_0(x) = \frac{\delta Z[J]}{\delta J(x)}\biggr|_{J=0}.
\end{equation}

We can follow the changing of the regulator with a (functional) differential equation (Wetterich-equation, or FRG-equation \cite{Gies_2012}). In the Euclidean case the Wetterich equation for the effective action reads
\begin{equation}
  \d_k\Gamma_k = \frac12 \hat\d_k \Tr\ln\left(\Gamma_k^{(2)} +
    R_k\right),
\end{equation}
where $\hat\d_k$ acts on the $k$ variable of the regulator only\footnote{That is $\hat\d_k f[R_k,\Gamma_k] = \left.\d_{k'} f[R_{k'},\Gamma_k]\right|_{k'=k}$.}. Clearly the Wetterich equation is sensible only for the domain where the second derivative exists.

To solve the FRG equations one usually assumes an Ansatz. One of the simplest choice is the local potential approximation (LPA) where the $\Gamma^{(n)}$ proper vertices are assumed to be local for $n>2$, and we also neglect the wave function renormalization term. In the scalar model which will be our example theory, the Ansatz reads:
\begin{equation}
  \label{eq:Gamma}
  \Gamma[\Phi] = \int d^dx\,\left[\frac12 (\d_\mu\Phi)(\d^\mu\Phi) + U(\Phi)\right],
\end{equation}
Applying Litim's optimized regulator \cite{Litim_2001} we arrive at the equation for the evolution of the potential
\begin{equation}
    \label{eq:runUk0}
    \d_k U_k = U_0 + \frac{\Omega_d k^{d+1}}{k^2 + \d_\Phi^2U_k},\qquad \Omega_d^{-1} = (4\pi)^{d/2} \Gamma(\frac d2+1),
\end{equation}
where the last $\Gamma$ is the gamma-function. To have numerically simpler form, we rescale $U_k = \Omega_d V_k$ and $\Phi = \sqrt{\Omega_d}\varphi$ to obtain
\begin{equation}
    \label{eq:runUk}
    \d_k V_k = \frac{k^{d+1}}{k^2 + \d_\varphi^2V_k}.
\end{equation}

\section{Effective potential around the vacuum}
\label{sec:vacuumexp}

As it was discussed in the Introduction, we want to establish an FRG evolution for the effective potential, expanded around the true vacuum situated at $\varphi=0$. To accommodate to the changing position of the minimum (the changing vacuum), we need to apply a slight modification to the usual FRG equations (c.f. \cite{Kobayashi_2020}). We shall write
\begin{equation}
    V_{k+dk}(\varphi- dk \delta A_k) = V_k(\varphi) + dk RHS,
\end{equation}
where $\delta A_k$ is a $k$-dependent constant, and ${RHS}$ is the right hand side of \eqref{eq:runUk}. This means
\begin{equation}
    \partial_k V_k = \delta A_k \partial_\varphi V_k + \frac{k^{d+1}}{k^2 +\d_\varphi^2V_k}.
\end{equation}
The task of the $\delta A_k$ term is to ensure that the minimum stays at $\varphi=0$. This means that we have to ensure
\begin{equation}
    \label{eq:nolinear}
    \partial_\varphi V_k(\varphi=0) = 0.
\end{equation}

We also know that the effective potential is convex, at least in a small neighbourhood of the vacuum. This means in particular
\begin{equation}
    \label{eq:posmass}
    \partial_\varphi^2 V_k(\varphi=0) \ge 0.
\end{equation}
To avoid IR divergences, we use an approach where the curvature is strictly positive (although we can be arbitrarily close to zero).

Our model in this paper is the scalar model up to the renormalizable $\Phi^4$ term. As we also discussed in the Introduction, the effective potential is not symmetric in the broken phase. We shall write therefore the LPA Ansatz as
\begin{equation}
    \label{eq:Gammapotential}
    V_k(\varphi) = \frac{m_k^2}2 \varphi^2 +\frac{g_k}6 \varphi^3 + \frac{\lambda_k}{24}\varphi^4.
\end{equation}
Note that the linear term is missing, according to \eqref{eq:nolinear}. If $g_k=0$, then $\Gamma[\varphi] = \Gamma[-\varphi]$, so the action is $\mathbb Z_2$ symmetric; if $g\neq0$ then the symmetry is broken. From the condition \eqref{eq:posmass} and the discussion after it, we will have $m_k^2>0$.

In \eqref{eq:Gammapotential} there is no difference between the spontaneous and explicit symmetry breaking. The spontaneity of the symmetry breaking is reflected by the Ward identity
\begin{equation}
    \label{eq:Ward}
    g_k^2 = 3\lambda_k m_k^2.
\end{equation}
A formal derivation of this statement using the background field method  can be found in the next section. But in the FRG evolution we shall recover this relation, if the evolution leads from a symmetric regime to the broken phase regime. This is in fact a consistency requirement of the FRG evolution equations, c.f. \cite{Gies_2012}.

To solve \eqref{eq:runUk}, we have to take into account that the right hand side is not a polynomial, so we have to apply an expansion, and truncate the resulting expression. We will use the following recipe:
\begin{equation}
    \label{eq:trunc}
    \frac{1}{k^2 +\d_\varphi^2V_k} \approx \frac1{\omega^2} - \frac{\partial_\varphi^2 V_k -m_k^2}{\omega^4} + \frac{(\partial_\varphi^2 V_k -m_k^2)^2}{\omega^6} + \dots,
\end{equation}
where $m_k^2=\partial^2_\varphi V(0)$ and $\omega^2 = k^2+m_k^2$, and we omit all further terms. The justification of this truncation scheme is that, as it will turn out soon, it is compatible with the Ward-identity \eqref{eq:Ward}.

With this truncation we can determine $\delta A_k$ from the condition that there is no linear term
\begin{equation}
    \delta A_k =k^{d+1}  \frac{g_k}{\omega^4 m_k^2}.
\end{equation}
Then we obtain
\begin{eqnarray}
  \label{eq:Phi4run}
  && \d_k m_k^2 = \frac{k^{d+1}} {\omega^4} \left( -\lambda +\frac{g_k^2}{m_k^2} (1+\frac{2m_k^2}{\omega^2}) \right),\nn
  && \d_k g_k = \frac{k^{d+1}g_k\lambda_k} {m_k^2\omega^4} \left[1 + \frac{6m_k^2}{\omega^2}\right],\nn
  && \d_k \lambda_k = \frac{6 k^{d+1}\lambda_k^2} {\omega^6}.
\end{eqnarray}

These are the equations that we propose to describe both the symmetric and the SSB phase scale evolution.

\section{Alternative derivation of the evolution equations}
\label{sec:vacuumexp_alt}

In this section we give an alternative derivation for equations \eqref{eq:Phi4run}, using the standard technique based on the evolution of the field expectation value. After we know the value of the background field, we can expand the potential around the actual minimum.

In the symmetric phase we find
\begin{equation}
    V_k = \frac12 m_k^2 \varphi^2 + \frac1{24}\lambda_k \varphi^4,
\end{equation}
and so
\begin{equation}
    \partial_\varphi^2 V_k-m_k^2 = \frac12\lambda_k\varphi^2.
\end{equation}
We also find
\begin{equation}
    \partial_k V_k = \frac12 \partial_k m_k^2 \varphi^2 + \frac1{24}\partial_k\lambda_k \varphi^4.
\end{equation}
Using \eqref{eq:trunc} we obtain
\begin{equation}
    \label{eq:runsym}
    \partial_k m_k^2 = -\frac{k^{d+1}\lambda_k}{\omega_k^4},\qquad
    \partial_k\lambda_k = \frac{k^{d+1}6\lambda_k^2}{\omega_k^6}.
\end{equation}
These equations are the same as \eqref{eq:Phi4run} for $g=0$.

In the broken phase we have a background field $\Phi_k$, characterizing the position of the minimum of the potential:
\begin{equation}
    V_k = \frac{\lambda_k}{24} (\varphi^2-\Phi_k^2)^2.
\end{equation}
We expand the potential around the $\varphi=\Phi_k$ minimum to find
\begin{equation}
    V_k = \frac{\lambda_k}6 \Phi_k^2\varphi^2 + \frac{\lambda_k}6 \Phi_k\varphi^3 + \frac{\lambda_k}{24} \varphi^4.
\end{equation}
The curvature mass an the cubic coupling read therefore
\begin{equation}
    \label{eq:mkgSSB}
    m_k^2=\frac{\lambda_k}3 \Phi_k^2,\quad g_k = \lambda_k\Phi_k.
\end{equation}
From these equations the the Ward identity \eqref{eq:Ward} follows immediately.

We obtain from \eqref{eq:mkgSSB}
\begin{equation}
    \partial_\varphi^2 V_k-m_k^2 = \frac{\lambda_k}2 (\varphi^2-\Phi_k^2) = \lambda_k\varrho,
\end{equation}
where $\varrho = (\varphi^2-\Phi_k^2)/2$. The $k$ derivative of the potential reads
\begin{equation}
    \partial_k V_k = \frac16\partial_k \varrho^2 -\frac{\lambda_k}6 (\partial_k\Phi_k^2)\varrho.
\end{equation}
Then from \eqref{eq:trunc} we find
\begin{equation}
    \partial_k\Phi_k^2=\frac{6k^{d+1}}{\omega_k^4}, \qquad \partial_k\lambda_k = \frac{k^{d+1}6\lambda_k^2}{\omega_k^6}.
\end{equation}
The equation for the curvature mass reads:
\begin{equation}
    \label{eq:MrunSSB}
    \partial_k m_k^2 = \frac13\partial_k(\lambda_k\Phi_k^2) = \frac{k^{d+1}2\lambda_k}{\omega_k^4} + \frac{k^{d+1} 6\lambda_k m_k^2}{\omega_k^6}.
\end{equation}
Using the Ward identity \eqref{eq:Ward}, we find agreement with the equations \eqref{eq:Phi4run}.

We can collect the symmetric and the broken phase equations introducing the "Ward ratio"
\begin{equation}
    \label{eq:Wardratio}
    r_k^2 = \frac{g_k^2}{3\lambda_k m_k^2} = \left\{
    \begin{array}{cc}
         0 &\quad\mathrm{symmetric}  \\
         1 &\quad{SSB}.
    \end{array}
    \right.
\end{equation}
Then \eqref{eq:runsym} and \eqref{eq:MrunSSB} can be unified as
\begin{equation}
    \partial_k m_k^2 = \frac{k^{d+1}}{\omega_k^4} \left(-\lambda_k(1-r_k^2) + \lambda_k r_k^2 (2 + \frac{6m_k^2}{\omega_k^2}) \right).
\end{equation}
Using the actual form of $r_k^2$ we find
\begin{equation}
    \partial_k m_k^2 = \frac{k^{d+1}}{\omega_k^4} \left(-\lambda_k + \frac{g_k^2}{m_k^2} (1 + \frac{2m_k^2}{\omega_k^2}) \right).
\end{equation}
This reproduces the first equation of \eqref{eq:Phi4run}.

We can also derive the evolution equation for the cubic coupling. In the symmetric phase it is zero, in the broken phase it is
\begin{equation}
    \partial_k g^2 = r_k^2 \partial_k(\lambda_k^2\Phi_k^2) = r_k^2\left(\frac{k^{d+1} 12 \lambda_k^3\Phi_k^2}{\omega_k^6} + \frac{k^{d+1} 6\lambda_k^2}{\omega_k^4}\right).
\end{equation}
By substituting $\lambda_k\Phi_k^2=3m_k^2$ we obtain:
\begin{equation}
    \partial_k g^2 = \frac{k^{d+1} 2\lambda_k g_k^2}{\omega_k^4}\left(1 + \frac{6m_k^2}{\omega_k^2}\right).
\end{equation}
This reproduces the second equation of \eqref{eq:Phi4run}.

\section{Results}
\label{sec:results}

Before we solve the equations numerically, we shall analyze them qualitatively. First of all, since the derivative of $g_k$ is proportional to $g_k$, thus $g_k\equiv0$ seems to be a good solution -- at least for $m_k^2>0$. At $m_k^2=0$, namely, the $g_k^2/m_k^2$ factors in \eqref{eq:Phi4run} yield $0/0$ expressions. And we can also realize that at $g_k=0$ the derivative of $m_k^2$ is negative, so sooner or later we encounter this weird situation.

\begin{figure}
    \centering
    \includegraphics[height=5cm]{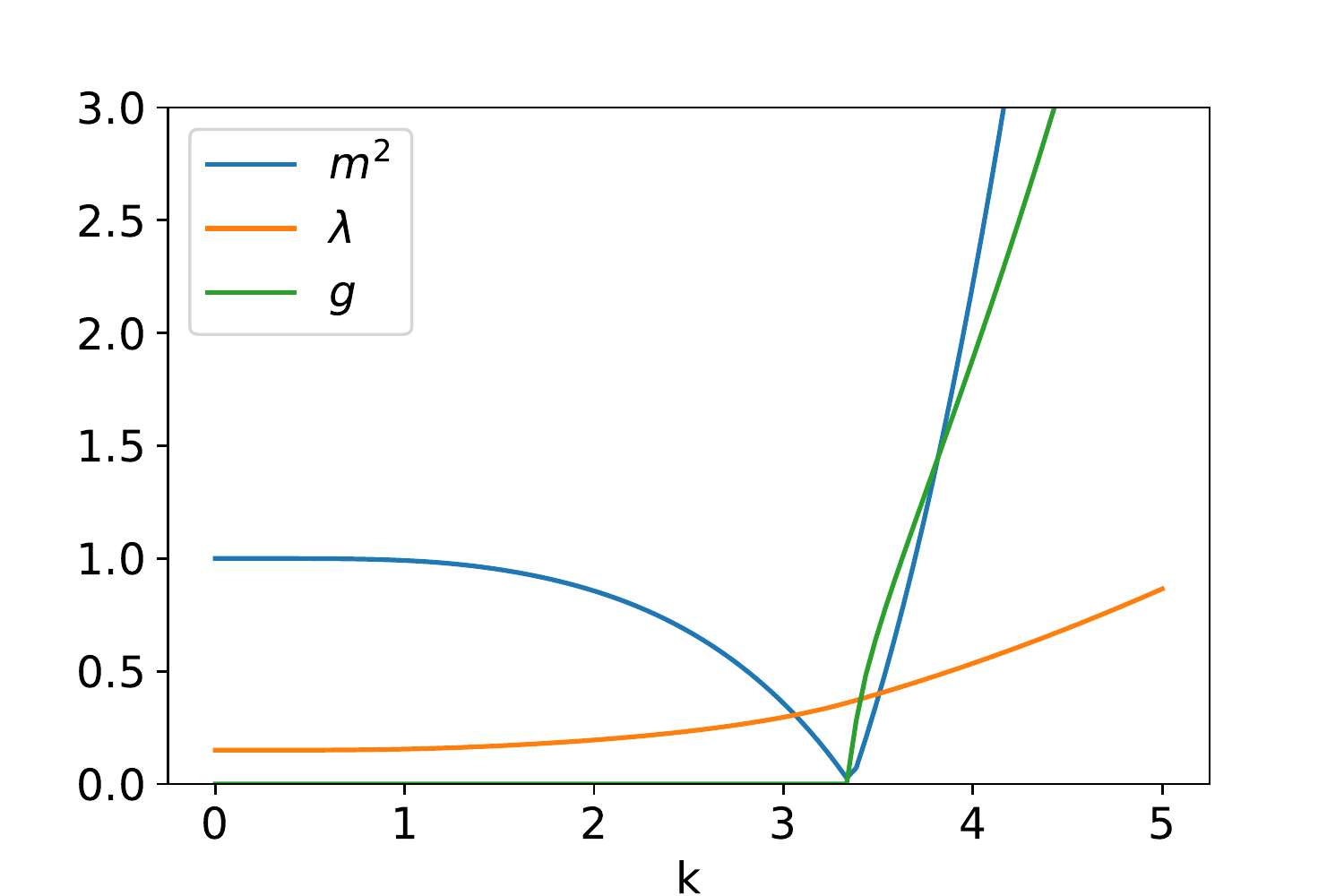}
    \caption{Running of the coupling near the SSB point with coupling values $(m_0^2, g_0, \lambda_0) = (1,10^{-6},0.15)$ at $k=0$.}
    \label{fig:running_near_SSB}
\end{figure}

To avoid the singular behaviour we shall start the running of $g_k$ with a nonzero value. The numerical solution provides the plot shown in Figure~\ref{fig:running_near_SSB}.
As we expected, the mass decreases, and tries to cross zero. But when it decreases below $g_k$, the $g_k^2/m_k^2$ term starts to dominate, and it overcompensates the $-\lambda_k$ term in the running of the mass. Consequently, the mass start to grow again, an we get into the broken phase.

\begin{figure}
    \centering
    \includegraphics[height=5cm]{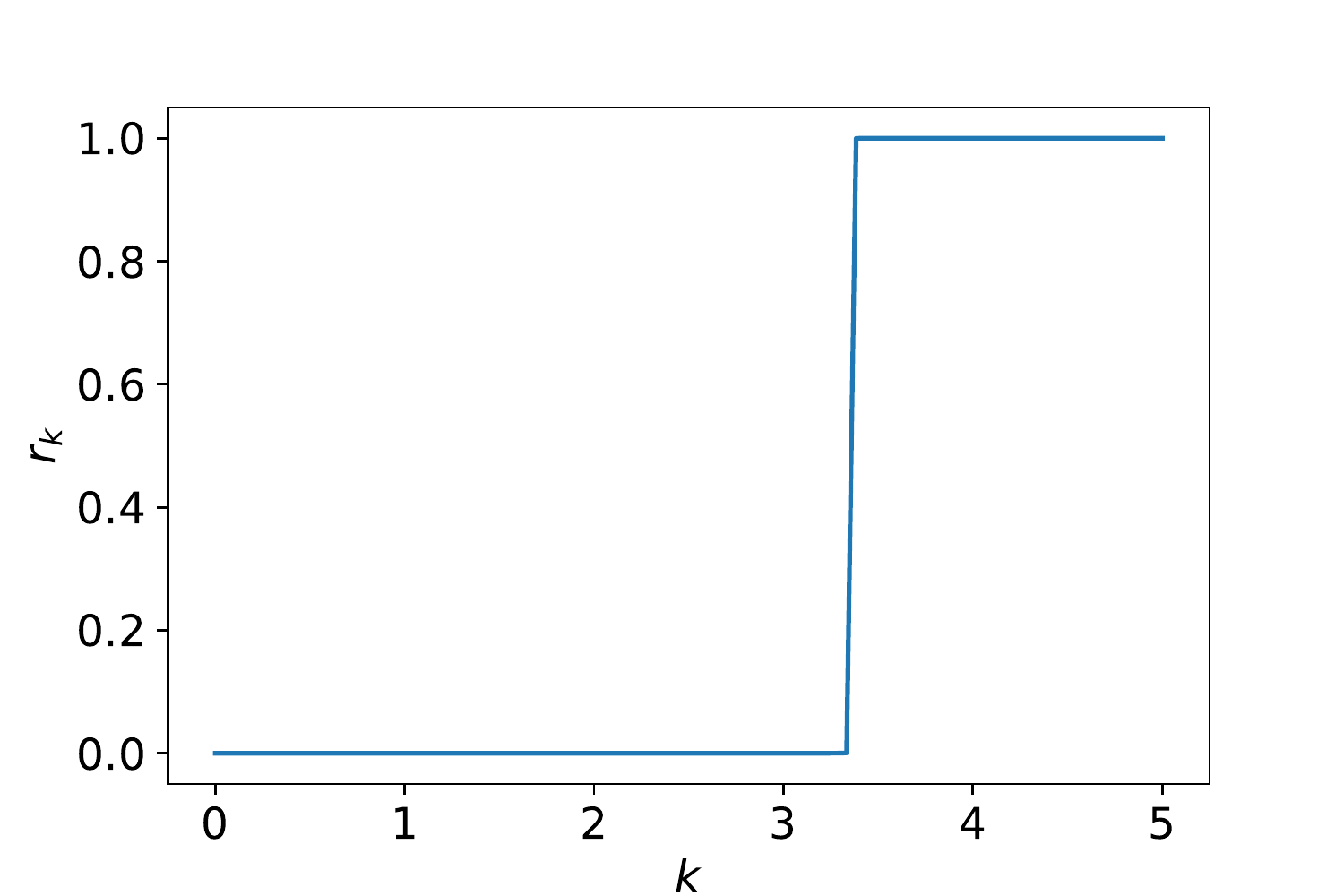}
    \caption{This figure shows the fulfillment of the Ward identity in a run with coupling values $(m_0^2, g_0, \lambda_0) = (1,10^{-6},0.15)$ at $k=0$. In symmetric phase we expect $r_k=0$, in the broken phase $r_k=1$. Numerical solution support these expectations.}
    \label{fig:Ward}
\end{figure}

We can also check, how well the Ward identities are satisfied. Therefore we plot the Ward ratio of \eqref{eq:Wardratio} in Figure~\ref{fig:Ward}.
As we see, the Ward ratio almost immediately takes the $r_k^2=1$ value characteristic for the SSB phase.

To understand the symmetry breaking better, we derive an equation for the Ward ratio:
\begin{equation}
    \partial_k r_k = \frac{k^{d+1}}{\omega_k^4} \frac{3\lambda_k}{2m_k^2} \left(1+\frac{2m_k^2}{\omega_k^2}\right) r_k(1-r_k^2).
\end{equation}
As we can see, the right hand side is zero at $r_k=\pm1$ and at $r_k=0$. These are "partial fixed points", meaning that although the running has no fixed point, but a special combination of them is still constant. The $r_k=\pm1$ are UV stable, which means that if we increase the scale we run into this partial fixed point. The $r_k=0$ is IR stable, we approach this partial fixed point by decreasing the scale. The speed of change is proportional to $1/m_k^2$, which means that in the vicinity of the phase transition point the change in the Ward ratio is very fast.

We shall also emphasize that the physics at $r_k=1$ and at $r_k=-1$ are very similar. The two fixed points differ only in the sign of the cubic coupling, and all correlation functions are the same up to a sign. Therefore the $\mathbb Z_2$ symmetry is not manifested in the dynamics, but it is manifested in the partial fixed point structure of the Ward ratio. This statement is the analog of the standard wisdom, saying that the symmetry is manifest in the dynamics, and it is broken by the background field value.

A very interesting feature of the above equations is the mass term appearing in the numerator. This means that the process of approaching a fixed point is accelerated considerably near $m_k^2=0$ points. This is an example of the IR divergences in the FRG equations. 

\section{Discussion and conclusions}
\label{sec:discussion}

The formalism derived in this paper uses exclusively the parameters of the Lagrangian (effective action) to describe SSB. Here we  avoided all of the potentially problematic consequences of the traditional treatment using Landau theory, like the proliferation of the vacua, or the flattening of the effective potential. For our formalism to go through, the effective potential need to be defined only in the infinitesimal neighborhood of the physical vacuum, which domain is gauge invariant in the gauge theories.

In this scenario the symmetry is broken by an explicit term in the action. In the symmetric phase the symmetry breaking terms are zero, in the SSB phase they are nonzero. In the $\Phi^4$ model studied in this paper, the only symmetry breaking renormalizable term was the cubic $\Phi^3$ with a coupling $g$. Here $g=0$ corresponds the symmetric, $g\neq 0$ the broken phase. The $\mathbb Z_2$ symmetry is manifested in the fact that for $g\to-g$ the correlation functions remain the same.

It is very interesting, how Ward identities appear in the system. If we start from a symmetric phase, and the symmetry breaking occurs through the scale evolution, then in the broken phase the Ward identities are automatically satisfied to a good precision. We can also study the dynamics of the Ward identity by following the scale evolution of the Ward ratio $r_k$ (c.f. \eqref{eq:Wardratio}). The evolution equation shows partial fixed points, where the Ward ratio is zero or $\pm1$. These partial fixed points can be used to characterize the phase of the system, and as so they can substitute the order parameter. The symmetry is either realized in the spectrum, or it is realized in the partial fixed point structure, transforming $r=+1$ to $r=-1$.

We still have a handle to speak about "field expectation value" using only the proper vertices, since
\begin{equation}
    \Phi_k = \frac{g_k}{\lambda_k},
\end{equation}
and both $g_k$ and $\lambda_k$ are well defined.

A very interesting property of the proposed equations is the accelerated running around the scale where the phase transition occurs, i.e. when $m_k^2\approx 0$. This is an "IR divergence" in the FRG running equations, since the running is coming from the correlation of large volumes. This property assures also that the Ward identity, characteristic for spontaneously symmetry breaking, will be satisfied immediately after a phase transition, if we started from a $g\approx 0$ action. This is true despite the fact that the formalism is not symmetric, only the initial state. From this point of view a symmetry breaking is "spontaneous", if it comes from an initial condition that is (almost) symmetric.

A possible consequence of this technique is that it can give an account for a "classical" process, the formation of the "classical field" purely based on notions that are sensible in quantum theory. By generalizing this thoughts we may get a better understanding of the measurement theory where a quantum wave starts to behave as a classical point mass.

\section*{Acknowledgment}
The authors acknowledge helpful discussions with G. Fej\H{o}s, A. Patk\'os. This work has been supported by the Hungarian Research Fund NKFIH (OTKA) under contract No. K123815.

\bibliographystyle{ieeetr}
\bibliography{refs}

\end{document}